\documentclass[aps,twocolumn]{revtex4-1}

\usepackage[dvipdfmx]{graphicx}
\usepackage{type1cm}
\usepackage[usenames,svgnames]{xcolor}
\usepackage{color}
\usepackage{url}
\usepackage{natbib}
\usepackage[dvipdfmx]{hyperref}
\hypersetup{
     colorlinks=true,       		
     linkcolor=Navy,          	
     citecolor=Navy,            
     filecolor=Navy,      		
     urlcolor=Navy,           	
    runcolor=cyan,
 }

\usepackage{amsmath,amssymb,bm,amsthm}

\def\beq{\begin{equation}}
\def\eeq{\end{equation}}
\def\nbeq{\begin{equation*}}
\def\neeq{\end{equation*}}

\def\<{\langle}
\def\>{\rangle}

\renewcommand{\d}{\partial}

\newtheorem{theorem}{Theorem}

\begin{document}
\title{Floquet prethermalization in periodically driven classical spin systems}

\author{Takashi Mori}
\affiliation{
Department of Physics, Graduate School of Science,
University of Tokyo, Bunkyo-ku, Tokyo 113-0033, Japan
}

\begin{abstract}
It is proved that the energy absorption in a periodically driven classical spin system is exponentially slow in frequency, which results in a two-step relaxation called the Floquet prethermalization.
This result is shown by establishing the classical limit of the quantum spin dynamics.
The Floquet prethermal state is well described by the Gibbs ensemble with respect to the static effective Hamiltonian obtained by a truncation of the Floquet-Magnus expansion.
On the other hand, the same effective Hamiltonian does not reproduce the local dynamics for an exponentially long time.
This is due to the chaoticity of classical dynamics, and in stark contrast to quantum spin systems, in which the effective Hamiltonian well reproduces the exact quantum dynamics up to an exponentially long time. 
\end{abstract}
\maketitle

\section{Introduction}
Periodically driven systems thermally isolated from the environment exhibit remarkable properties that are not easily realized in equilibrium systems.
For instance, dynamical localization~\cite{Dunlap1986,Grifoni1998,Kayanuma2008}, coherent destruction of tunneling~\cite{Grossmann1991,Grifoni1998,Kayanuma2008}, and quantum phase transitions induced by periodic driving~\cite{Eckardt2005,Eckardt2009,Zenesini2009} are remarkable nonequilibrium phenomena.
Recent experimental advances also triggered studies of Floquet topological states both experimentally~\cite{Aidelsburger2013, Atala2013, Jotzu2014, Aidelsburger2015} and theoretically~\cite{Kitagawa2010, Lindner2011, Kitagawa2011_transport, Dora2012, Delplace2013, Grushin2014, Sentef2015, Titum2016, Takasan2017, Hubener2017}.
The Floquet time crystal is also a recent hot topic~\cite{Else2016,Yao2017,Zhang2017}. 
Floquet engineering aims to design such novel states of matter by periodic driving.

One of the strategies of Floquet engineering in quantum systems is to consider in the regime of high frequency of the driving field.
It is now recognized that a meaningful effective static Hamiltonian is obtained in the high-frequency regime by using the Floquet-Magnus expansion~\cite{Maricq1982,Mori2015_Floquet}.
Although it is believed that a non-integrable periodically driven system ultimately heats up to infinite temperature~\cite{Lazarides2014,DAlessio2014,Ponte2015}, the effective static Hamiltonian describes quasi-stationary states appearing in an intermediate time scale that grows exponentially with respect to frequency~\cite{Kuwahara2016,Mori2016_rigorous,Abanin2017_effective,Abanin2017_rigorous}.

The relaxation to a quasi-stationary state before reaching the true stationary state is called prethermalization.
See Ref.~\cite{Mori_review2018} for a recent review on thermalization and prethermalization in isolated quantum systems.
Prethermalization under a fast driving is a generic feature of periodically driven quantum lattice systems, and it is called Floquet prethermalization.
Emergence of the long intermediate time scale stems from the fact that the energy absorption due to periodic driving is exponentially slow in generic many-body quantum spin systems
\footnote
{This statement also holds for lattice fermion systems, but there is no rigorous proof for generic lattice boson systems yet because of unboundedness of boson creation and annihilation operators}.
A Floquet prethermal state in a periodically driven quantum system is described by the effective static Hamiltonian obtained by a truncation of the Floquet-Magnus expansion.

Recently, it has been numerically shown that \textit{classical} systems under fast periodic driving also have exponentially long intermediate time scales in which the system stays in quasi-stationary states~\cite{Rajak_arXiv2018,Howell_arXiv2018}.
Similarly to the quantum systems, a quasi-stationary state in a classical spin system is described by the effective static Hamiltonian obtained by a truncation of the Floquet-Magnus expansion~\cite{Howell_arXiv2018} (see Sec.~\ref{sec:discussion} for the definition of the Floquet-Magnus expansion in generic classical systems).
So far, there is no rigorous justification for using such an effective static Hamiltonian in classical systems.

The purpose of this paper is to prove that this is justified in periodically driven classical spin systems, which implies that Floquet prethermalization also occurs for general classical spin systems.
We first show that the classical dynamics of a many spin system is obtained by the large-$S$ limit of the Schr\"odinger equation under a quantum spin-$S$ Hamiltonian.
Next, we show that a rigorous theorem on the exponentially slow energy absorption in quantum spin systems, which has been previously proved in Refs.~\cite{Kuwahara2016,Mori2016_rigorous}, is applicable to a quantum spin-$S$ Hamiltonian for an arbitrary $S$.
This means that classical spin systems behave in a similar way as quantum spin systems; the energy absorption is exponentially slow and the system exhibits Floquet prethermalization.
In contrast, it turns out that another rigorous theorem in quantum spin systems, which states that the local dynamics is well reproduced by the effective static Hamiltonian obtained by a truncation of the Floquet-Magnus expansion~\cite{Kuwahara2016}, does not hold in classical spin systems.
We argue that this is due to the chaoticity of classical dynamics.

This paper is organized as follows.
In Sec.~\ref{sec:FM}, we introduce the Floquet-Magnus expansion in quantum systems and summarize the previously known two rigorous results on it.
In Sec.~\ref{sec:large-S}, we formulate a classical spin system as the large-$S$ limit of a quantum spin-$S$ system.
We rigorously prove that the large-$S$ limit of the quantum dynamics starting from a factorized initial state is reduced to the corresponding classical dynamics.
In Sec.~\ref{sec:FM_classical}, we show that the rigorous result on the exponentially slow energy absorption presented in Sec.~\ref{sec:FM} is applicable to a quantum spin-$S$ system for an arbitrarily large $S$, while the other theorem on dynamics of local quantities is not.
This result tells us that the Floquet prethermalization also occurs in classical spin systems, but microscopic classical dynamics is not accurately reproduced by the effective static Hamiltonian.
We discuss our result and an open problem in Sec.~\ref{sec:discussion}.

\section{Floquet-Magnus expansion in quantum spin systems}
\label{sec:FM}

We aim to establish the presence of Floquet prethermalization in classical spin systems.
Since a classical spin system can be regarded as the large-$S$ limit of the corresponding quantum spin-$S$ system, we can utilize several properties of quantum spin systems to predict behavior of a classical spin system.

In this section, we summarize important properties of the Floquet-Magnus expansion in periodically driven quantum spin systems.

Let us consider generic quantum lattice systems on a $d$-dimensional lattice.
The set of sites is denoted by $\Lambda$.
The total number of sites is given by $|\Lambda|=N$ (for a set $A$, $|A|$ denotes the number of elements in $A$).
Each site $i\in\Lambda$ at position $\bm{r}_i\in\mathbb{R}^d$ has its own Hilbert space $\mathcal{H}_i$.
The distance $d(i,j)$ between two sites $i$ and $j$ is defined by $d(i,j)=|\bm{r}_i-\bm{r}_j|$.
For $X\subset\Lambda$, we define $\mathcal{H}_X:=\bigotimes_{i\in X}\mathcal{H}_i$.
The Hilbert space of the total system is thus written as $\mathcal{H}_{\Lambda}$.
The Hamiltonian $\hat{H}(t)$ with time $t$ satisfies $\hat{H}(t)=\hat{H}(t+T)$, where $T$ is the period of the driving.

We assume that the Hamiltonian satisfies the \textit{$k$-locality}~\cite{Kuwahara2016}; $\hat{H}(t)$ is written in the form
\beq
\hat{H}(t)=\sum_{X\subset\Lambda: |X|\leq k}\hat{h}_X(t)
\label{eq:k-local}
\eeq
for an integer $k$, where $\hat{h}_X(t)$ is an operator acting nontrivially only to $\mathcal{H}_X$.
The physical meaning of the $k$-locality is that the Hamiltonian contains up to $k$-site mutual interactions.
It is noted that interactions may be long-ranged ($k$-locality does not necessarily imply the spatial locality).

An important parameter $g>0$ is introduced as an arbitrary constant that satisfies
\beq
g\geq \sup_{t\in[0,T]}\max_{i\in\Lambda}\sum_{X\subset\Lambda: i\in X}\|\hat{h}_X(t)\|,
\label{eq:g-extensive}
\eeq
where the operator norm is denoted by $\|\cdot\|$.
Intuitively, $g$ represents the possible largest value of the single-site energy.
In a normal quantum spin system with the extensive energy, $g$ is independent of the system size $N$.

The time evolution over a single period from $t=0$ to $t=T$ is called the Floquet operator, which is given by \beq
\hat{U}_{\mathrm F}:=\mathcal{T}e^{-i\int_0^Tdt\,\hat{H}(t)},
\eeq
where $\mathcal{T}$ denotes the time-ordering operator.
Throughout the paper, we set $\hbar=1$.
The Floquet Hamiltonian $\hat{H}_{\mathrm F}$ is defined by
\beq
\hat{U}_{\mathrm F}=e^{-i\hat{H}_{\mathrm F}T}.
\eeq
At stroboscopic times $t=MT$ with $M$ being an integer, the quantum state $|\Psi(t)\>$ evolving with the Schr\"odinger equation $id|\Psi(t)\>/dt=\hat{H}(t)|\Psi(t)\>$ is expressed as
\beq
|\Psi(t)\>=\hat{U}_{\mathrm F}^M|\Psi(0)\>=e^{-i\hat{H}_Ft}|\Psi(0)\>.
\eeq

This expression implies that the Floquet Hamiltonian plays the role of an effective static Hamiltonian, but it is in general not straightforward to obtain $\hat{H}_{\mathrm F}$ from $\hat{H}(t)$.
When the frequency of the driving is large, the Floquet-Magnus expansion is known as a systematic high-frequency expansion of $\hat{H}_{\mathrm F}$:
\beq
\hat{H}_{\mathrm F}=\sum_{m=0}^{\infty}T^m\hat{\Omega}_m,
\label{eq:expansion}
\eeq
where the explicit form of $\hat{\Omega}_n$ is given by~\cite{Bialynicki-Biula1969}
\begin{widetext}
\begin{align}
\hat{\Omega}_n&=&
\sum_{\sigma}
{ (-1)^{n-\theta[\sigma]}\theta[\sigma]!(n-\theta[\sigma])! \over 
i^n(n+1)^2n! T^{n+1} }
\int_0^Tdt_1\int_0^{t_1}dt_2
\dots\int_0^{t_n}dt_{n+1}
 [\hat{H}(t_{\sigma(1)}),[\hat{H}(t_{\sigma(2)}),\dots,[H(t_{\sigma(n)}),H(t_{\sigma(n+1)})]\dots]],
\label{eq:FM}
\end{align}
\end{widetext}
where $\sigma$ is a permutation and $\theta[\sigma]=\sum_{i=1}^n\theta(\sigma(i+1)-\sigma(i))$ with $\theta(\cdot)$ is the step function.
For instance, the two lowest terms are given by
\beq
\left\{
\begin{split}
&\hat{\Omega}_0=\frac{1}{T}\int_0^Tdt_1\,H(t_1), \\
&\hat{\Omega}_1=\frac{1}{2iT^2}\int_0^Tdt_1\int_0^{t_1}dt_2\,[H(t_1),H(t_2)].
\end{split}
\right.
\eeq

When $T$ is small, it is expected that we can approximately truncate the Floquet-Magnus expansion as
\beq
\hat{H}_{\mathrm F}\approx\sum_{m=0}^nT^m\hat{\Omega}_m=:\hat{H}_{\mathrm F}^{(n)}.
\label{eq:truncation}
\eeq
When $\hat{H}(t)$ is a $k$-local Hamiltonian, $\hat{H}_{\mathrm F}^{(n)}$ is at most a $(nk)$-local Hamiltonian.
Therefore, if the truncation (\ref{eq:truncation}) provides us a good approximation, it would imply that the system has a (quasi-)local conserved quantity that is very close to $\hat{H}_{\mathrm F}^{(n)}$.
Moreover, if $\hat{H}_{\mathrm F}^{(n)}$ is a non-integrable Hamiltonian, it is also expected that $\hat{H}_{\mathrm F}^{(n)}$ obeys the eigenstate thermalization hypothesis~\cite{Deutsch1991,Srednicki1994,Mori_review2018} (ETH) and the system relaxes to a stationary state described by the Floquet-Gibbs state
\beq
\rho_{\mathrm{FG}}^{(n)}:=\frac{e^{-\beta\hat{H}_{\mathrm F}^{(n)}}}{\mathrm{Tr}\,e^{-\beta\hat{H}_{\mathrm F}^{(n)}}},
\label{eq:FG}
\eeq
where the inverse temperature $\beta$ is determined from the condition $\<\psi(0)|\hat{H}_{\mathrm F}^{(n)}|\psi(0)\>=\mathrm{Tr}\,\hat{H}_{\mathrm F}^{(n)}\rho_{\mathrm{FG}}^{(n)}$.

However, it is generally believed that the exact Floquet Hamiltonian $\hat{H}_{\mathrm F}$ obeys the \textit{Floquet ETH}, for any local operator $\hat{O}$ and any pair of eigenstates $|u_k\>$ and $|u_l\>$ of $\hat{H}_{\mathrm F}$,
\beq
\<u_k|\hat{O}|u_k\>\approx\<u_l|\hat{O}|u_l\>.
\label{eq:Floquet-ETH}
\eeq
this means that every Floquet eigenstate $|u_k\>$ is locally indistinguishable from the infinite-temperature state:
\beq
\<u_k|\hat{O}|u_k\>\approx\mathrm{Tr}\,\hat{O}\frac{\hat{1}_{\Lambda}}{D_{\Lambda}},
\eeq
where $\hat{1}_{\Lambda}$ is the identity operator acting onto $\mathcal{H}_{\Lambda}$, and $D_{\Lambda}:=\dim\mathcal{H}_{\Lambda}$.
The Floquet ETH implies that the system relaxes to the stationary state described by the infinite-temperature ensemble $\hat{1}_{\Lambda}/D_{\Lambda}$: for any local operator $\hat{O}$,
\beq
\<\psi(t)|\hat{O}|\psi(t)\>\approx\mathrm{Tr}\,\hat{O}\frac{\hat{1}_{\Lambda}}{D_{\Lambda}}
\eeq
for sufficiently large $t$~\cite{Lazarides2014,DAlessio2014,Ponte2015}.
It is noted that the Floquet-Gibbs state $\rho_{\mathrm{FG}}^{(n)}$ with $\beta>0$ locally differs from the infinite-temperature state $\hat{1}_{\Lambda}/D_{\Lambda}$.
Therefore, if $\hat{H}_{\mathrm F}$ is replaced by a truncated one $\hat{H}_{\mathrm F}^{(n)}$ in the time evolution, it contradicts the prediction of the Floquet ETH.

Recent works~\cite{Kuwahara2016,Mori2016_rigorous,Abanin2017_effective,Abanin2017_rigorous} have shown that this contradiction is resolved by the fact that the Floquet-Magnus expansion~(\ref{eq:expansion}) is not a convergent series in general and a finite-temperature Floquet-Gibbs state actually describes a quasi-stationary state appealing in an intermediate time scale before reaching the infinite-temperature state~\footnote
{
It is believed that the convergence radius of Eq.~(\ref{eq:expansion}) tends to zero in the thermodynamic limit for generic nonintegrable systems.
The Floquet ETH implies the divergence of the Floquet-Magnus expansion, but the latter does not necessarily imply the former. 
}.
The truncated Floquet Hamiltonian $\hat{H}_{\mathrm F}^{(n)}$ is not an approximation of a strict conserved quantity $\hat{H}_{\mathrm F}$, but $\<\psi(t)|\hat{H}_{\mathrm F}^{(n)}|\psi(t)\>$ at stroboscopic times changes exponentially slowly in the high-frequency regime.

This property is precisely described by the following theorem~\cite{Kuwahara2016,Mori2016_rigorous}:
\begin{theorem}
\label{theorem1}
Assume $8kgT\leq 1$.
At stroboscopic times $t=MT$ with $M$ an integer, the truncated Flouqet Hamiltonian $\hat{H}_{\mathrm F}^{(n)}$ with $n\leq n_0:=\lfloor 1/(8kgT)-1\rfloor$ satisfies
\begin{align}
\frac{1}{N}\left|\<\psi(t)|\hat{H}_{\mathrm F}^{(n)}|\psi(t)\>-\<\psi(0)|\hat{H}_{\mathrm F}^{(n)}|\psi(0)\>\right|
\\
\leq 16kg^22^{-n_0}t+C_nT^{n+1}
\end{align}
for any initial state $|\psi(0)\>\in\mathcal{H}_{\Lambda}$, where $|\psi(t)\>=e^{-i\hat{H}_{\mathrm F}t}|\psi(0)\>$ is the quantum state at time $t$, and $C_n$ is a positive constant depending only on $n$, $k$, and $g$.
\end{theorem}

For $n=0$, $\hat{H}_{\mathrm F}^{(0)}=\hat{\Omega}_0=(1/T)\int_0^Tdt\,\hat{H}(t)$ is the time-averaged Hamiltonian, which is interpreted as the energy of the system.
Then, Theorem~\ref{theorem1} for $n=0$ implies that the energy absorption due to periodic driving is exponentially slow with respect to the frequency $\omega=2\pi/T$ (note that $2^{-n_0}=e^{-O(\omega)}$).

For a short-range interacting system, we can show a stronger result.
The Hamiltonian (\ref{eq:k-local}) is said to be short-ranged if
\beq
\max_{i\in\Lambda}\sup_{t\in[0,T]}\sum_{X\subset\Lambda: i\in X, \mathrm{diam}(X)\geq r}\|h_X(t)\|\leq F(r)
\label{eq:LR_condition}
\eeq
for all $r>0$, where $\mathrm{diam}(X):=\max_{i,j\in X}d(i,j)$ and $F(r)$ is a function how interactions decay with the distance $r$.
Here we assume exponentially decaying interactions $F(r)\sim e^{-\kappa r}$ with some constant $\kappa>0$.
In this case, we can prove the Lieb-Robinson bound for arbitrary local operators $\hat{O}_X$ and $\hat{O}_Y$ acting nontrivially onto $X\subset\Lambda$ and $Y\subset\Lambda$, respectively:
\beq
\left\|[\hat{O}_X,\hat{O}_Y]\right\|\leq ce^{-(\ell-vt)/\xi}\min(|X|,|Y|)\|\hat{O}_X\|\cdot\|\hat{O}_Y\|,
\label{eq:LR}
\eeq
where $\ell=d(X,Y):=\min_{i\in X,j\in Y}d(i,j)$, and $c$, $v$, and $\xi$ are positive constants depending on $k$, $g$, and $F(r)$.

By using the Lieb-Robinson bound, we obtain the following theorem~\cite{Kuwahara2016}:
\begin{theorem}
\label{theorem2}
Assume short-range interactions in a $d$-dimensional regular lattice and $16kgT\leq 1$.
At stroboscopic times $t=MT$ with $M$ an integer,
\begin{align}
\left|\<\psi(t)|\hat{O}_X|\psi(t)\>-\<\psi^{(n_0')}(t)|\hat{O}_X|\psi^{(n_0')}(t)\>\right|
\nonumber \\
\leq\left(12g2^{-n_0'/2}+\frac{2c}{T}e^{(-\ell_0-vt)/\xi}\right)\|\hat{O}_X\|\cdot |X|t
\label{eq:theorem2}
\end{align}
for any initial state $|\psi(0)\>\in\mathcal{H}_{\Lambda}$ and any local operator $\hat{O}_X$ acting nontrivially onto $X\subset\Lambda$.
Here, $|\psi(t)\>=e^{-i\hat{H}_{\mathrm F}t}|\psi(0)\>$, $|\psi^{(n_0')}(t)\>=e^{-i\hat{H}_{\mathrm F}^{(n_0')}t}|\psi(0)\>$, $n_0'=\lfloor 1/(16kgT)-1\rfloor$, and $\ell_0=\mathrm{const.}\times 2^{n_0'/(2d)}=e^{O(\omega)}$.
\end{theorem}

It is noted that the right-hand side of Eq.~(\ref{eq:theorem2}) is small for any $t<e^{O(\omega)}$.
It means that the local dynamics of the system is well approximated by the Schr\"odinger equation under an effective static Hamiltonian $\hat{H}_{\mathrm F}^{(n_0')}$ up to an exponentially long time.
The truncated Floquet Hamiltonian is not just a nearly conserved quantity; it also governs the time evolution of local quantities up to a prethermal regime.

Later it turns out that, in the classical limit, Theorem~\ref{theorem1} still holds, which explains the Floquet prethermalization in classical spins~\cite{Howell_arXiv2018}, but Theorem~\ref{theorem2} does \textit{not} hold.

\section{Classical spins as the large-$S$ limit of quantum spins}
\label{sec:large-S}

It is a ``well-known'' fact that a quantum spin-$S$ system becomes classical in the limit of $S\rightarrow\infty$.
For equilibrium states, this is proved by Lieb~\cite{Lieb1973}.
For dynamics, the precise statement and its rigorous proof are not found in the literature, so we present them in this section and Appendix~\ref{appendix}.
We consider the following Hamiltonian of $N$ spin-$S$ system:
\beq
H(t)=-\frac{1}{2S}\sum_{ij}^N\sum_{\alpha,\beta=x,y,z}J_{ij}^{\alpha\beta}(t)\hat{S}_i^{\alpha}\hat{S}_j^{\beta} -\sum_{i=1}^N\bm{h}_i(t)\cdot\hat{\bm{S}}_i,
\label{eq:H}
\eeq
where $J_{ij}^{\alpha\beta}(t)$ and $\bm{h}_i(t)$ are time-dependent two-spin interactions and the local magnetic field, respectively.
Each lattice site is labeled by $i$, which has its own spin $\hat{\bm{S}}_i$ with $\hat{\bm{S}}_i^2=S(S+1)$ and $[\hat{S}_i^{\alpha},\hat{S}_j^{\beta}]=i\delta_{ij}\sum_{\gamma=x,y,z}\epsilon_{\alpha\beta\gamma}\hat{S}_i^{\gamma}$.
The interaction $J_{ij}^{\alpha\beta}(t)=J_{ji}^{\beta\alpha}(t)$ is arbitrary as long as it satisfies
\beq
\sum_{\alpha,\beta=x,y,z}\sum_j|J_{ij}^{\alpha\beta}(t)|\leq J_0
\label{eq:interaction}
\eeq
for any $i$ and all $t$ with some constant $J_0>0$ independent of $N$, $S$, and $\omega$.

We consider the quantum dynamics generated by Eq.~(\ref{eq:H}) with an initial state $|\Psi(0)\>$ and consider its large-$S$ limit.
What we want to prove here is that we can regard $\{\hat{\bm{S}}_i\}$ as continuous classical vectors in the large-$S$ limit.
We explain the precise meaning below.
Let us consider a factorized initial state $|\Psi(0)\>=\bigotimes_{i=1}^N|\psi_i(0)\>$, where $|\psi_i(0)\>$ is a state vector in $\mathcal{H}_i$.
We assume that the initial state is \textit{classical}, that is, if we define $\bm{S}_i(0):=\<\psi_i(0)|\hat{\bm{S}}_i|\psi_i(0)\>$, then $\bm{S}_i^2=S^2$ (the maximum value).
The state vector evolves as $id|\Psi(t)\>/dt=\hat{H}(t)|\Psi(t)\>$, and we write $\bm{S}_i(t)=\<\Psi(t)|\hat{\bm{S}}_i|\Psi(t)\>$.
Then, in the large-$S$ limit, the normalized spin vector $\bm{s}_i(t)=\bm{S}_i(t)/S$ obeys the following classical equations of motion:
\beq
\left\{
\begin{aligned}
&\frac{d}{dt}\bm{s}_i(t)=\bm{s}_i(t)\times\tilde{\bm{h}}_i(t), \\
&\tilde{h}_i^{\alpha}(t)=h_i^{\alpha}(t)+\sum_{j=1}^N\sum_{\beta=x,y,z}J_{ij}^{\alpha\beta}s_j^{\beta}(t).
\end{aligned}
\right.
\label{eq:EOM}
\eeq
Moreover, any correlation function is given by a product of the corresponding classical spin variables, i.e.,
\begin{align}
\lim_{S\rightarrow\infty}\frac{1}{S^n}\<\Psi(t)|\hat{S}_{i_1}^{\alpha_1}\hat{S}_{i_2}^{\alpha_2}\dots\hat{S}_{i_n}^{\alpha_n}|\Psi(t)\>
\nonumber \\
=s_{i_1}^{\alpha_1}(t)s_{i_2}^{\alpha_2}(t)\dots s_{i_n}^{\alpha_n}(t).
\label{eq:statement}
\end{align}
This is a precise meaning of the statement that quantum spins dynamically behave as classical in the large-$S$ limit.

The proof of this statement is provided in Appendix~\ref{appendix}.
For the proof, we consider the spin-1/2 decomposition of a spin-$S$ operator, which is introduced in the next section.
We remark that the result does not change if the Hamiltonian contains general $k$-body interactions with $k\geq 3$ of the form
$$\frac{1}{S^{k-1}}\sum_{i_1,i_2,\dots,i_k}^N\sum_{\alpha_1,\dots,\alpha_k=x,y,z}J_{i_1,\dots,i_k}^{\alpha_1,\dots,\alpha_k}(t)\hat{S}_{i_1}^{\alpha_1}\hat{S}_{i_2}^{\alpha_2}\dots \hat{S}_{i_k}^{\alpha_k},$$
as long as 
$$\sum_{\alpha_1,\dots,\alpha_k}\sum_{i_2,\dots,i_k}|J_{i_1,i_2,\dots,i_k}^{\alpha_1,\dots,\alpha_k}(t)|<J_0$$
for any $i_1$ with some constant $J_0$, although we assume two-body interactions in the proof in Appendix~\ref{appendix} for simplicity.

\section{Rigorous results on the Floquet-Magnus expansion in periodically driven classical spin systems}
\label{sec:FM_classical}

We discuss whether rigorous results on quantum spin systems, that is, Theorem~\ref{theorem1} and Theorem~\ref{theorem2},  are applicable to classical spins.
We have argued that a classical spin system is regarded as a quantum spin system in the large-$S$ limit.
Therefore, we consider applicability of the theorems for the Hamiltonian (\ref{eq:H}) in the large-$S$ limit.

The condition of applicability of Theorem~\ref{theorem1} is that the Hamiltonian is written in the form of Eq.~(\ref{eq:k-local}) and local operators $\hat{h}_X(t)$ satisfy Eq.~(\ref{eq:g-extensive}) for some $k$ and $g$ which are independent of $N$, $S$, and $\omega$.
In addition to it, for applicability of Theorem~\ref{theorem2}, interactions should be short-ranged in the sense that the inequality~(\ref{eq:LR_condition}) with $F(r)\sim e^{-\kappa r}$ is satisfied for all $r>0$.

By choosing $\hat{h}_X(t)=-(1/S)\sum_{\alpha\beta}J_{ij}^{\alpha\beta}(t)\hat{S}_i^{\alpha}\hat{S}_j^{\beta}$ for $X=\{i,j\}$ and $\hat{h}_X(t)=-\bm{h}_i(t)\cdot\hat{\bm{S}}_i$ for $X=\{i\}$, Eq.~(\ref{eq:H}) is expressed in the form of Eq.~(\ref{eq:k-local}) with $k=2$, but we have
\begin{align}
\sum_{X\subset\Lambda:i\in X}\|\hat{h}_X(t)\|=\sum_{j=1}^N\frac{1}{S}\left|\sum_{\alpha,\beta=x,y,z}J_{ij}^{\alpha\beta}(t)\hat{S}_i^{\alpha}\hat{S}_j^{\beta}\right|
\nonumber \\
+\left|\bm{h}_i(t)\cdot\hat{\bm{S}}_i\right|,
\end{align}
which is a quantity of $O(S)$ for large $S$.
It means that $g$ in the inequality (\ref{eq:g-extensive}) diverges in the large-$S$ limit.
Since Theorems~\ref{theorem1} and \ref{theorem2} are meaningful only for finite $g$, these theorems are naively not applicable.

\begin{figure}[t]
\begin{center}
\includegraphics[clip,width=8cm]{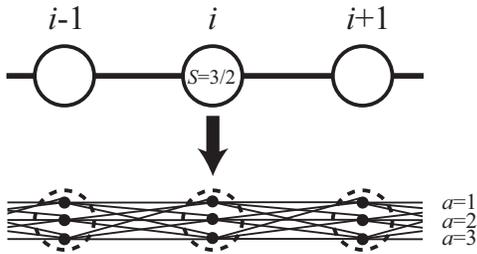}
\caption{A schematic picture of the decomposition of spin $S$ into spin-1/2s in a one-dimensional system.
This figure shows the case of $S=3/2$, and $a\in\{1,2,3\}$.}
\label{fig:decomp}
\end{center}
\end{figure}

Despite this naive consideration, we can make $g$ finite even in the large-$S$ limit.
We shall decompose each spin-$S$ operator $\hat{\bm{S}}_i$ into $2S$ spin-1/2 operators $\{\hat{s}_{i,a}\}$ with $a=1,2,\dots,2S$ as
\beq
\hat{\bm{S}}_i=\sum_{a=1}^{2S}\hat{\bm{s}}_{i,a},
\label{eq:decomp}
\eeq
where the Hilbert space must be restricted to the subspace with the maximum total spin, $\hat{\bm{S}}_i^2=S(S+1)$.
The Hamiltonian is written by
\beq
\hat{H}(t)=-\frac{1}{2S}\sum_{(i,a),(j,b)}\sum_{\alpha,\beta}J_{ij}^{\alpha\beta}(t)\hat{s}_{i,a}^{\alpha}\hat{s}_{j,b}^{\beta}-\sum_{(i,a)}\bm{h}_i\cdot\hat{\bm{s}}_{i,a}.
\label{eq:H_decomp}
\eeq
See Fig.~\ref{fig:decomp} for a schematic picture of the decomposition.

As explained in Sec.~\ref{sec:large-S}, an initial state should be factorized; $|\Psi(0)\>=\bigotimes_{i=1}^N|\psi_i(0)\>$.
The decomposition of Eq.~(\ref{eq:decomp}) with a restriction to the subspace with the maximum total spin yields for each $i$,
\beq
|\psi_i(0)\>=\bigotimes_{a=1}^{2S}|\phi_i(0)\>,
\eeq
where $|\phi_i(0)\>$ is a state vector in the two-dimensional Hilbert space representing a spin-1/2.
It is noted that $|\phi_i(0)\>$ does not depend on $a$, which means that all of $2S$ spins at site $i$ are in the same state.

We can suppose that each pair $(i,a)$ with $i=1,2,\dots,N$ and $a=1,2,\dots,2S$ defines a single site.
The Hamiltonian (\ref{eq:H_decomp}) is then written in the form of Eq.~(\ref{eq:k-local}) with $k=2$ by identifying
\beq
\hat{h}_X(t)=\left\{
\begin{alignedat}{2}
&-\frac{1}{S}\sum_{\alpha,\beta=x,y,z}&&J_{ij}^{\alpha\beta}(t)\hat{s}_{i,a}^{\alpha}\hat{s}_{j,b}^{\beta}\\
& & &\text{ for } X=\{(i,a),(j,b)\}, \\
&-\bm{h}_i(t)\cdot\hat{\bm{s}}_{i,a} & &\text{ for } X=\{(i,a)\}.
\end{alignedat}
\right.
\eeq
We have
\begin{align}
&\sum_{X\subset\Lambda: (i,a)\in X}\|\hat{h}_X(t)\|
\nonumber \\
&\leq\frac{1}{S}\sum_{j=1}^N\sum_{b=1}^{2S}\sum_{\alpha,\beta}\left|J_{ij}^{\alpha\beta}(t)\right|\|\hat{s}_{i,a}^{\alpha}\|\cdot\|\hat{s}_{j,b}^{\beta}\|+|\bm{h}_i(t)\cdot\hat{\bm{s}}_{i,a}|
\nonumber \\
&\leq\frac{1}{2}\sum_{j=1}^N\sum_{\alpha,\beta=x,y,z}\left|J_{ij}^{\alpha\beta}(t)\right|+\frac{1}{2}|\bm{h}_i(t)|
\nonumber \\
&\leq\frac{J_0+h_0}{2},
\end{align}
where we have used the inequality~(\ref{eq:interaction}) and have defined
\beq
h_0:=\max_{i\in\Lambda}\sup_{t\in[0,T]}|\bm{h}_i(t)|
\eeq
in the last line.
We assume that $h_0$ is independent of $N$, $S$, and $\omega$.

Thus, we can choose
\beq
g=\frac{J_0+h_0}{2},
\label{eq:g_quantum}
\eeq
which is finite in the limit of $S\rightarrow\infty$.

In this way, by expressing a spin-$S$ Hamiltonian as a spin-1/2 Hamiltonian (\ref{eq:H_decomp}) restricted to the subspace of the maximum spin of the resultant spin $\sum_{a=1}^{2S}\hat{\bm{s}}_{i,a}$, it is found that the condition of applicability of Theorem~\ref{theorem1} is satisfied.
Therefore, periodically driven classical spin systems also exhibit exponentially slow heating in the high-frequency regime, and hence a recent numerical observation~\cite{Howell_arXiv2018} mentioned in introduction is reasonably understood.

On the other hand, the applicability of Theorem~\ref{theorem2} is nontrivial.
In our quantum Hamiltonian~(\ref{eq:H_decomp}), each site $i$ is decomposed into sites $(i,a)$ with $a=1,2,\dots,2S$.
It means that we have an extra dimension, and there are long-range interactions along this extra dimension (see Fig.~\ref{fig:decomp}).
However, the proof of Theorem~\ref{theorem2} relies on the Lieb-Robinson bound, which requires short-range interactions~\footnote
{If the distance between the sites $(i,a)$ and $(j,b)$, which is denoted by $d[(i,a),(j,b)]$, is defined by the distance $d(i,j)$ between $i$ and $j$ in the original lattice before the decomposition, interactions in Eq.~(\ref{eq:H_decomp}) are short-ranged in the sense of Eq.~(\ref{eq:LR_condition}) and thus the Lieb-Robinson bound~(\ref{eq:LR}) holds.
Nevertheless, the original proof of Theorem~\ref{theorem2} is \textit{not} applicable due to the fact that the number of lattice points $(j,b)$ satisfying $d[(i,a),(j,b)]\leq \ell$ for some $\ell>0$ is roughly proportional to $\ell^dS$, which diverges in the classical limit.}.

Here, we argue that \textit{Theorem~\ref{theorem2} does not hold in chaotic classical spin systems}.
When $16kgT\leq 1$, we can show that
\beq
\left|\<\psi(T)|\hat{O}_X|\psi(T)\>-\<\psi^{(n_0')}(T)|\hat{O}_X|\psi^{(n_0')}(T)\>\right|\leq e^{-O(\omega)},
\label{eq:single_n0}
\eeq
whose proof is given in Ref.~\cite{Kuwahara2016}, and this inequality holds even in the classical limit since short-range interactions are not assumed in the proof.
This inequality tells us that $\hat{H}_\mathrm{F}^{(n_0')}$ well approximates the dynamics at least over a single period within an exponentially small error in $\omega$.
In classical chaotic systems, however, this small error will grow exponentially fast, and hence, in the classical limit, $\hat{H}_\mathrm{F}^{(n_0')}$ will give a good approximation of the dynamics up to the time proportional to $\omega$, which increases as $\omega$ but not exponentially.
This means that Theorem~\ref{theorem2} is not applicable to chaotic classical spin systems.

Because the calculation of $\hat{H}_\mathrm{F}^{(n_0')}$ for a concrete system is difficult (remember that $n_0'\propto\omega$ is very large), we consider a modified version of Theorem~\ref{theorem2}.
By properly modifying the proof of Theorem~\ref{theorem2} given in Ref.~\cite{Kuwahara2016}, we can show that under the same condition of Theorem~\ref{theorem2}, for any local operator $\hat{O}_X$ with $\|\hat{O}_X\|=1$ and large $t=MT$ ($M$ is an integer),
\beq
\left|\<\psi(t)|\hat{O}_X|\psi(t)\>-\<\psi^{(n)}(t)|\hat{O}_X|\psi^{(n)}(t)\>\right|\lesssim |X|T^{n+1}t^{d+1}
\label{eq:theorem2'}
\eeq
in a $d$-dimensional system, where $|\psi^{(n)}(t)\>=e^{-i\hat{H}_\mathrm{F}^{(n)}t}|\psi(0)\>$ and  $n$ is a nonnegative integer independent of $\omega$ ($n\ll n_0'$).
The proof of Eq.~(\ref{eq:theorem2'}) is given in Appendix~\ref{appendix:theorem2}.
The inequality~(\ref{eq:theorem2'}) tells us that the truncated Floquet Hamiltonian $\hat{H}_\mathrm{F}^{(n)}$ generates a good approximation of the exact dynamics up to the time
\beq
\tau^{(n)}\propto \omega^{(n+1)/(d+1)}.
\eeq
On the other hand, on the time evolution over a single period, without the assumption of short-range interactions, we can show
\beq
\left|\<\psi(T)|\hat{O}_X|\psi(T)\>-\<\psi^{(n)}(T)|\hat{O}_X|\psi^{(n)}(T)\>\right|\leq \alpha_n|X|T^{n+2}
\label{eq:single}
\eeq
for $n\ll n_0'$, where $|\psi^{(n)}(t)\>=e^{-iH_\mathrm{F}^{(n)}t}|\psi(0)\>$ and $\alpha_n$ is a constant that depends only on $n$, $k$ and $g$.
Equation~(\ref{eq:single}) can be proved by slightly modifying the proof of Eq.~(\ref{eq:single_n0}) provided in Ref.~\cite{Kuwahara2016}.
In classical chaotic systems, this small error will grow exponentially fast.
As a result, in the classical limit, the truncated Floquet Hamiltonian $\hat{H}_\mathrm{F}^{(n)}$ with $n\ll n_0$ will approximate the exact dynamics up to the time proportional $\ln\omega$, which is much shorter than $\tau^{(n)}$.

\begin{figure*}[tb]
\begin{minipage}[t]{0.42\hsize}
\begin{center}
\includegraphics[width=8cm]{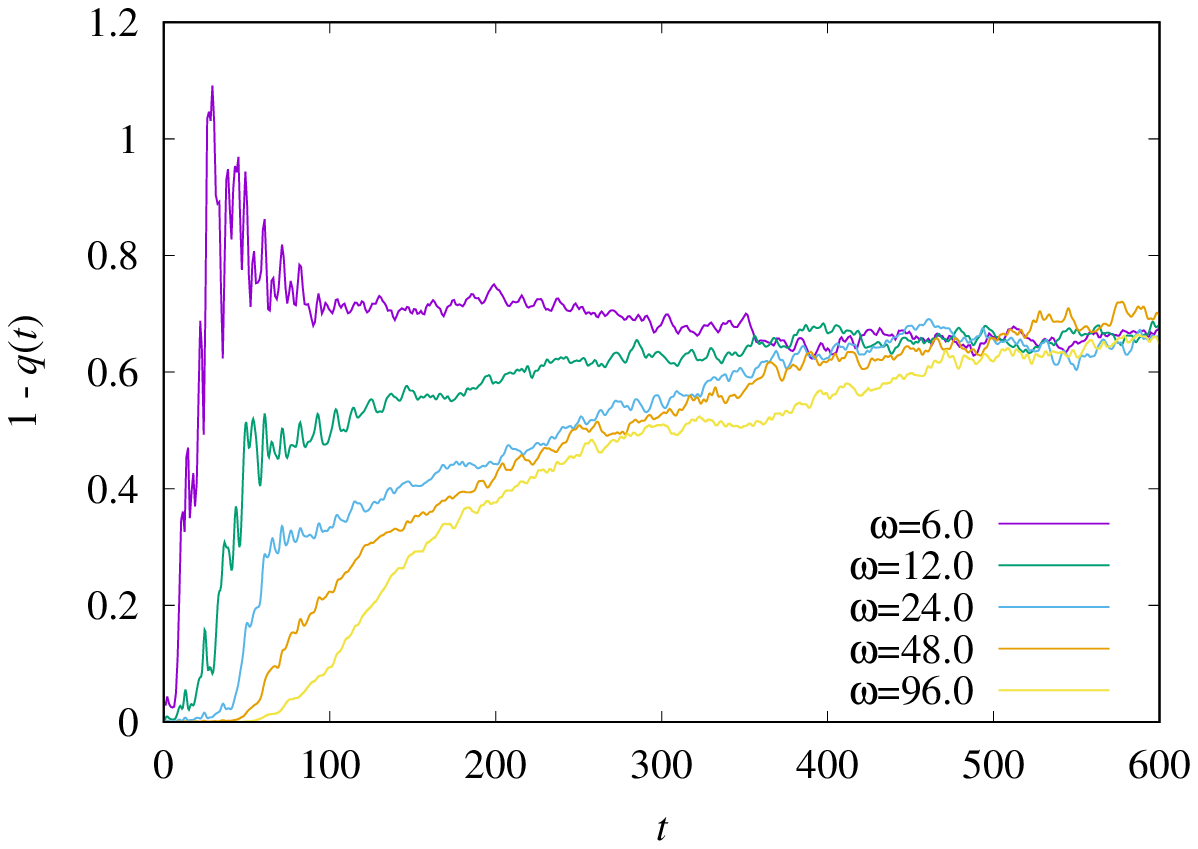}
\end{center}
\end{minipage}
\begin{minipage}[t]{0.42\hsize}
\begin{center}
\includegraphics[width=8cm]{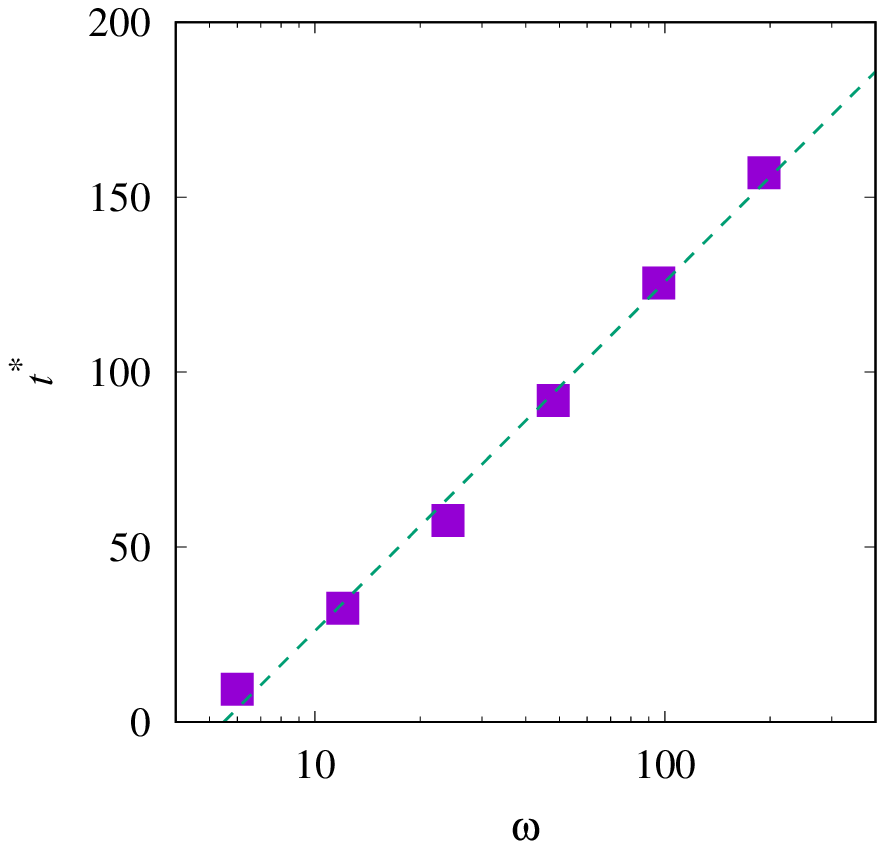}
\end{center}
\end{minipage}
\caption{Left: the time evolution of $1-q(t)$ for several values of $\omega$. Right: the $\omega$-dependence of $t^*$, which is defined by the time at which $1-q(t)$ exceeds 0.2.
It is found that $t^*\propto\ln\omega$, which indicates the violation of Theorem~\ref{theorem2}.}
\label{fig:overlap}
\end{figure*}

In order to confirm the violation of Theorem~\ref{theorem2} expressed by the inequality~(\ref{eq:theorem2'}), we numerically calculate the classical spin dynamics.
We consider the classical Hamiltonian of a one-dimensional spin system,
\beq
H(t)=\left\{
\begin{split}
&\sum_{i=1}^N(Js_i^zs_{i+1}^z-h_zs_i^z) & &\text{for }t\in[0,T/2] \bmod T, \\
&-\sum_{i=1}^Nh_xs_i^x & &\text{for }t\in(T/2,T)\bmod T.
\end{split}
\right.
\label{eq:num_H}
\eeq
We choose $J=1$, $h_x=-0.9045$, and $h_x=-0.809$.
The classical dynamics is given by Eq.~(\ref{eq:EOM}) with $\tilde{\bm{h}}_i(t)=-\d H(t)/\d\bm{s}_i$.
The initial state is randomly chosen as
\beq
s_i^x=\sin\phi_i, \quad s_i^y=0, \quad s_i^z=\cos\phi_i,
\eeq
where $\{\phi_i\}$ are iid random variables uniformly chosen from $[-\pi/100,\pi/100]$.
This model is same as the one studied in Ref.~\cite{Howell_arXiv2018}.

We now compare the exact dynamics with the approximate dynamics generated by 
\beq
H_\mathrm{F}^{(0)}=\frac{1}{2}\sum_{i=1}^N(Js_i^zs_{i+1}^z-h_zs_i^z-h_xs_i^x).
\eeq
Let us denote by $\{\bm{s}_i(t)\}$ and $\{\bm{s}_i'(t)\}$ the solution of the exact classical equations of motion and that of the equations of motion generated by $H_\mathrm{F}^{(0)}$, respectively, starting from the same initial state.
The difference between them is quantified by the overlap $q(t)\in[0,1]$ defined by
\beq
q(t)=\frac{1}{N}\sum_{i=1}^N\bm{s}_i(t)\cdot\bm{s}_i'(t).
\eeq
Two spin configurations $\{\bm{s}_i(t)\}$ and $\{\bm{s}_i'(t)\}$ are close to each other if $1-q(t)\ll 1$.
If the inequality~(\ref{eq:theorem2'}) holds, $1-q(t)$ should remain small up to the time $\tau^{(0)}\propto\omega^{1/2}$.
This dependence is different from $\ln\omega$ expected from Eq.~(\ref{eq:single}) and the chaoticity of classical dynamics.

The distance $1-q(t)$ averaged over 32 realizations of initial states is plotted as a function of $t$ for several values of $\omega$ in the left of Fig.~\ref{fig:overlap}.
We define $t^*$ as the minimum time at which $1-q(t)$ exceeds 0.2, and $\omega$-dependence of $t^*$ is given in the right of Fig.~\ref{fig:overlap}.
It turns out that $t^*$ grows only logarithmically in $\omega$, which is much shorter than $\tau^{(0)}\propto\omega^{1/2}$ and consistent with the above argument expected from classical chaoticity.
This result shows that the inequality (\ref{eq:theorem2'}) does not hold.
Since Theorem~\ref{theorem2} is proved under the same condition of the inequality~(\ref{eq:theorem2'}), this numerical result strongly supports the argument that Theorem~\ref{theorem2} does not hold in classical spin systems.

\begin{figure}
\centering
\includegraphics[width=7.5cm]{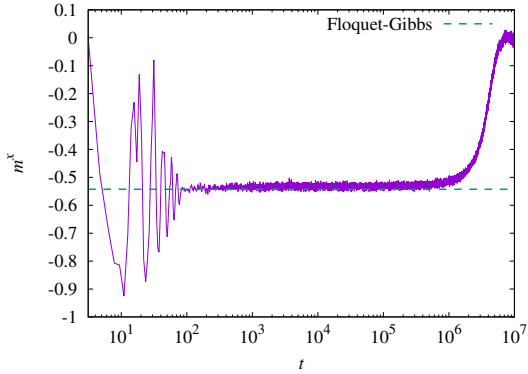}
\caption{The time evolution of $m^x(t)=(1/N)\sum_{i=1}^Ns_i^x(t)$ for $\omega=4.0$ (solid line) and the prethermal value predicted by the Floquet-Gibbs state with $n=0$ (dashed line).
After the initial relaxation, the system is in a quasi-stationary state described by the Floquet-Gibbs state.
In a much longer timescale, the energy absorption takes place and the system finally heats up to the infinite temperature.}
\label{fig:pre}
\end{figure}

It should be emphasized that the violation of the classical-spin counterpart of Theorem~\ref{theorem2} does not mean that the Floquet prethermal state is not described by the Floquet-Gibbs state given by Eq.~(\ref{eq:FG}).
Figure~\ref{fig:pre} shows long time evolutions of $m^x(t)=(1/N)\sum_{i=1}^Ns_i^x(t)$ for $\omega=4.0$ averaged over 32 realizations of initial states.
One can see that the Floquet-Gibbs state reproduce the prethermal values of $m^x(t)$, and the heating takes place in a much longer timescale compared to the initial relaxation.

\section{Discussion}
\label{sec:discussion}

The reason why we have succeeded in showing that the Floquet-Magnus expansion is relevant to classical spin systems stems from the fact that the classical limit (the large-$S$ limit) is well-controlled mathematically.
The present proof of the classical counterpart of Theorem~\ref{theorem1} cannot be extended to generic classical systems.

In interacting classical systems, the equations of motion are nonlinear and there is no Floquet theory~\cite{Bukov_review2015}.
However, the Floquet operator and the Floquet Hamiltonian for classical systems can be formally formulated by considering the classical Liouville equation
\beq
\frac{\d P(q,p,t)}{\d t}=\{H(t),P(q,p,t)\}=:-i\hat{L}(t)P(q,p,t),
\eeq
where $q$ and $p$ are the sets of positions and momenta of all the particles in the system, respectively, $P(q,p,t)$ is the distribution function in the phase space, and $\{\cdot,\cdot\}$ denotes the Poisson bracket.
The Liouville equation is structurally similar to the Schr\"odinger equation, and we can go along the same line as in the quantum case.
The time evolution operator over a single period is given by
\beq
\hat{U}_{\mathrm F}^{\mathrm{(cl)}}=\mathcal{T}e^{-i\int_0^Tdt\hat{L}(t)}=:e^{-i\hat{L}_{\mathrm F}T},
\eeq
and we can formally consider the Magnus expansion of $\hat{L}_{\mathrm F}$ as
\beq
\hat{L}_{\mathrm F}=\sum_{m=0}^{\infty}T^m\hat{\Xi}_m,
\label{eq:L_Magnus}
\eeq
where $\hat{\Xi}_m$ is given in the same way as in Eq.~(\ref{eq:FM}) ($\hat{H}(t)$ is just replaced by $\hat{L}(t)$).
For example,
\beq
\left\{
\begin{split}
&\hat{\Xi}_0=\frac{1}{T}\int_0^Tdt_1\,\hat{L}(t_1), \\
&\hat{\Xi}_1=\frac{1}{2iT^2}\int_0^Tdt_1\int_0^{t_1}dt_2[\hat{L}(t_1),\hat{L}(t_2)].
\end{split}
\right.
\eeq

Now we formally define the classical Floquet Hamiltonian by
\beq
-i\hat{L}_{\mathrm F}=\left\{H_{\mathrm F}^{\mathrm{(cl)}},\cdot\right\},
\eeq
and we shall derive the inverse-frequency expansion of $\hat{H}_{\mathrm F}^{\mathrm{(cl)}}$ based on the Floquet-Magnus expansion of $\hat{L}_{\mathrm F}$.
It is found that for two Liouville operators $\hat{L}_1=i\{H_1,\cdot\}$ and $\hat{L}_2=i\{H_2,\cdot\}$, the commutator $[\hat{L}_1,\hat{L}_2]=\hat{L}_1\hat{L}_2-\hat{L}_2\hat{L}_1$ is expressed by
\beq
[\hat{L}_1,\hat{L}_2]=i\left\{i\{H_1,H_2\},\cdot\right\}.
\eeq
This equation implies that taking the commutator of Liouville operators corresponds to taking the Poisson bracket of the Hamiltonians and multiplying the factor $i$.

Therefore, the classical Floquet Hamiltonian $H_{\mathrm F}^{\mathrm{(cl)}}$ is formally given by
\beq
H_{\mathrm F}^{\mathrm{(cl)}}=\sum_{m=0}^{\infty}T^m\Omega_m^{\mathrm{(cl)}},
\label{eq:FM_classical}
\eeq
where $\Omega_m^{\mathrm{(cl)}}$ is obtained by replacing $\hat{L}(t)$ and their commutators $[\hat{L}(t_j),\hat{L}(t_k)]$ in $\hat{\Xi}_m$ by $H(t)$ and $i\{H(t_j),H(t_k)\}$, respectively.
For example,
\beq
\left\{
\begin{split}
&\Omega_0^{\mathrm{(cl)}}=\frac{1}{T}\int_0^Tdt_1\,H(t_1), \\
&\Omega_1^{\mathrm{(cl)}}=\frac{1}{2T^2}\int_0^Tdt_1\int_0^{t_1}dt_2\{H(t_1),H(t_2)\},
\end{split}
\right.
\eeq
and so on.

One will realize that the inverse-frequency expansion of $H_{\mathrm F}^{\mathrm{(cl)}}$ in the classical case is obtained by formally replacing the commutators $(1/i)[\cdot,\cdot]$ appearing in the Floquet-Magnus expansion in the quantum case~(\ref{eq:FM}) by the Poisson brackets $\{\cdot,\cdot\}$.
An effective static Hamiltonian for a classical system is formally defined by a truncation of the expansion~(\ref{eq:FM_classical}).

The present paper has focused on classical spin systems and has shown that there exists the classical counterpart of the theorem derived for quantum systems; the effective static Hamiltonian is a quasi-conserved quantity, which explains the Floquet prethermalization in classical spin systems~\cite{Howell_arXiv2018}.
We emphasize that the Floquet prethermalization has also been observed numerically in classical many-body kicked rotors~\cite{Rajak_arXiv2018}.
It is an open problem to clarify the property of the effective Hamiltonian in more general classical systems.

\begin{acknowledgments}
The author thanks an anonymous referee for pointing out a fundamental flaw in the previous version on Theorem~\ref{theorem2} in classical spin systems.
This work was supported by JSPS KAKENHI Grants. No. 15K17718.
\end{acknowledgments}

\appendix
\section{Proof of the classical spin dynamics as the large-$S$ limit of the quantum spin dynamics}
\label{appendix}

We consider the spin-1/2 decomposition introduced in Sec.~\ref{sec:FM_classical}.
By introducing the notation
\beq
\hat{V}_{(i,a),(j,b)}(t):=\sum_{\alpha,\beta=x,y,z}J_{ij}^{\alpha\beta}(t)\hat{s}_{i,a}^{\alpha}\hat{s}_{j,b}^{\beta},
\eeq
the Hamiltonian~(\ref{eq:H_decomp}) is written as
\beq
\hat{H}(t)=\frac{1}{2S}\sum_{(i,a),(j,b)}\hat{V}_{(i,a),(j,b)}(t)-\sum_{(i,a)}\bm{h}_i(t)\cdot\hat{\bm{s}}_{i,a}.
\eeq

Let us consider the $n$-spin reduced density matrix 
\beq
\rho_{\Lambda_0}(t)\equiv {\rm Tr}_{\Lambda_0^c}|\Psi(t)\>\<\Psi(t)|,
\label{eq:reduced}
\eeq
where $\Lambda_0=\{(i_1,a_1),(i_2,a_2),\dots, (i_n,a_n)\}$ is a set of the sites ($i_k=1,2,\dots, N$ and $a_k=1,2,\dots, 2S$) with $(i_k,a_k)\neq (i_l,a_l)$ for any $k\neq l$, and $\Lambda_0^c$ is the complement of $\Lambda_0$, i.e., $\Lambda_0^c=\{(i,a)\in\Lambda\setminus\Lambda_0\}$.
The partial trace over the Hilbert space $\mathcal{H}_{\Lambda_0^c}$ is denoted by $\mathrm{Tr}_{\Lambda_0^c}$.

We introduce $\Lambda_1=\Lambda_0\cup(j_1,b_1)$ with some $(j_1,b_1)\in\Lambda_0^c$.
Similarly, we write
\beq
\Lambda_k=\Lambda_0\cup\{(j_l,b_l)\}_{l=1}^k
\eeq
with $(j_l,b_l)\in\Lambda_0^c$ and $(j_l,b_l)\neq (j_m,b_m)$ for $l\neq m$.
We will use the simple notation $$\sum_{\Lambda_k}=\sum_{(j_1,b_1),(j_2,b_2),\dots, (j_k,b_k)}.$$
The reduced density matrix on the set $\Lambda_k$ is defined in the same manner as Eq.~(\ref{eq:reduced}).

The equation of motion for $\rho_{\Lambda_k}(t)$ is given by
\begin{align}
\frac{d}{dt}\rho_{\Lambda_k}(t)=-i\mathcal{L}_{\Lambda_k}(t)\rho_{\Lambda_k}(t)&+\sum_{(j_{k+1},b_{k+1})}\mathcal{W}_{\Lambda_{k+1}}(t)\rho_{\Lambda_{k+1}}(t)
\nonumber \\
&+\frac{1}{S}\mathcal{V}_{\Lambda_k}(t)\rho_{\Lambda_k}(t),
\end{align}
where the super-operators $\mathcal{L}_{\Lambda_k}(t)$, $\mathcal{W}_{\Lambda_{k+1}}(t)$, and $\mathcal{V}_{\Lambda_k}(t)$ are defined by
\begin{align}
\mathcal{L}_{\Lambda_k}(t)\rho_{\Lambda_k}:=\left[-\sum_{(ia)\in\Lambda_k}\bm{h}_i(t)\cdot\hat{\bm{s}}_{ia},\rho_{\Lambda_k}\right],
\end{align}
\begin{align}
\mathcal{W}_{\Lambda_{k+1}}(t)&\rho_{\Lambda_{k+1}}:=-\frac{i}{S}\sum_{(ia)\in\Lambda_k}
\nonumber \\
&{\rm Tr}_{(j_{k+1}b_{k+1})}\left[\hat{V}_{(ia),(j_{k+1}b_{k+1})}(t),\rho_{\Lambda_{k+1}}\right],
\end{align}
and
\beq
\mathcal{V}_{\Lambda_k}(t)\rho_{\Lambda_k}:=-\frac{i}{2}\sum_{(ia),(jb)\in\Lambda_k}\left[\hat{V}_{(ia),(jb)}(t),\rho_{\Lambda_k}\right].
\eeq
We will also define the following unitary super-operator for later convenience:
\beq
U_{\Lambda_k}(t,t')=\mathcal{T}e^{-i\int_{t'}^tds\mathcal{L}_{\Lambda_k}(s)}
\eeq
for $t'\leq t$.

The classical equation of motion, Eq.~(\ref{eq:EOM}) is reproduced if the reduced density matrix is given by $\rho_{\Lambda_k}(t)=\sigma_{\Lambda_k}(t)$, where
\beq
\sigma_{\Lambda_k}(t)=\bigotimes_{(ia)\in\Lambda_k}|\phi_{ia}(t)\>\<\phi_{ia}(t)|
\eeq
with $|\phi_{ia}(t)\>=|\phi_{i1}(t)\>$ for any $a=1,2,\dots,2S$.
The state vector of the spin-1/2 on site $(i,a)$ obeys the following equation:
\beq
i\frac{d}{dt}|\phi_{ia}(t)\>=\hat{H}_{\rm cl}^{(ia)}(t)|\phi_{ia}(t)\>,
\label{eq:classical_ia}
\eeq
where
\beq
\hat{H}_{\rm cl}^{(ia)}=-\sum_{j=1}^N\sum_{\alpha,\beta=x,y,z}J_{ij}^{\alpha\beta}(t)s_j^{\beta}(t)\hat{s}_{ia}^{\alpha}-\bm{h}_i(t)\cdot\hat{\bm{s}}_{ia}
\eeq
with 
\beq
s_{j}^{\beta}(t)=\frac{1}{S}\sum_{b=1}^{2S}\<\phi_{jb}(t)|\hat{s}_{jb}^{\beta}|\phi_{jb}(t)\>,
\eeq
where it is noted that $\<\phi_{jb}(t)|\hat{s}_{jb}^\beta(t)|\phi_{jb}(t)\>$ is independent of $b$.
By using Eq.~(\ref{eq:classical_ia}), the following equation of motion for $\sigma_{\Lambda_k}(t)$ is derived:
\begin{align}
\frac{d}{dt}\sigma_{\Lambda_k}(t)=-i\mathcal{L}_{\Lambda_k}(t)\sigma_{\Lambda_k}(t)+\sum_{(j_{k+1},b_{k+1})}\mathcal{W}_{\Lambda_{k+1}}(t)\sigma_{\Lambda_{k+1}}(t)
\nonumber \\
+\frac{1}{S}\sum_{(j,b)\in\Lambda_k}\mathcal{V}'_{\Lambda_k,(jb')}\sigma_{\Lambda_k\cup(jb')}(t),
\label{eq:sigma}
\end{align}
where
\begin{align}
&\mathcal{V}'_{\Lambda_k,(jb')}(t)\sigma_{\Lambda_k\cup(jb')}(t)
\nonumber \\
&:=-\frac{i}{2}\sum_{(ia)\in\Lambda_k}{\rm Tr}_{(jb')}[\hat{V}_{(ia),(jb')}(t),\sigma_{\Lambda_k\cup(jb')}(t)].
\end{align}
Here, $b'\in\{1,2,\dots, 2S\}$ is an arbitrary integer satisfying $(j,b')\notin\Lambda_k$ (there is at least one such $b'$ for any $j$ as long as $n+k<2S$).

We can show Eqs.~(\ref{eq:EOM}) and (\ref{eq:statement}) if $\|\rho_{\Lambda_0}(t)-\sigma_{\Lambda_0}(t)\|_{1}^{\Lambda_0}\rightarrow 0$ in the limit of $S\rightarrow+\infty$.
Here, the trace norm of an operator $\hat{A}_{\Lambda_k}$ on the Hilbert space $\mathcal{H}_{\Lambda_k}$ is denoted by $\|\hat{A}_{\Lambda_k}\|_{1}^{\Lambda_k}:={\rm Tr}_{\Lambda_k}\sqrt{\hat{A}^{\dagger}_{\Lambda_k}\hat{A}_{\Lambda_k}}$.
It is noted that the trace norm of the density matrix is unity, 
\beq
\|\rho_{\Lambda_k}(t)\|_1^{\Lambda_k}=\|\sigma_{\Lambda_k}(t)\|_1^{\Lambda_k}=1.
\eeq

Similarly, the operator norm of $\hat{A}_{\Lambda_k}$ is denoted by 
\beq
\|\hat{A}_{\Lambda_k}\|^{\Lambda_k}:=\sup_{\Psi\in\mathcal{H}_{\Lambda_k},\<\Psi|\Psi\>=1}\sqrt{\<\Psi|\hat{A}^{\dagger}_{\Lambda_k}A_{\Lambda_k}|\Psi\>}.
\eeq
We will use the following well-known inequalities:
\begin{align}
\|\hat{A}_{\Lambda_k}\hat{B}_{\Lambda_k}\|_1^{\Lambda_k}\leq\|\hat{A}_{\Lambda_k}\|^{\Lambda_k}\|\hat{B}_{\Lambda_k}\|_1^{\Lambda_k}, \\
\|\hat{A}_{\Lambda_k}+\hat{B}_{\Lambda_k}\|_1^{\Lambda_k}\leq\|\hat{A}_{\Lambda_k}\|_1^{\Lambda_k}+\|\hat{B}_{\Lambda_k}\|_1^{\Lambda_k}.
\end{align}

By performing the Duhamel expansion~\cite{Mori2013_exactness}, $\rho_{\Lambda_0}(t)$ and $\sigma_{\Lambda_0}(t)$ are expressed as follows:
\begin{widetext}
\begin{align}
\rho_{\Lambda_0}(t)=\sum_{l=0}^L\sum_{\Lambda_l}\int_0^tdt_1\int_0^{t_1}dt_2\dots\int_0^{t_{l-1}}dt_l\, U_{\Lambda_0}(t,t_1)\mathcal{W}_{\Lambda_1}(t_1)U_{\Lambda_1}(t_1,t_2)\mathcal{W}_{\Lambda_2}(t_2)
\nonumber \\
\dots U_{\Lambda_{l-1}}(t_{l-1},t_l)\mathcal{W}_{\Lambda_l}(t_l)U_{\Lambda_l}(t_l,0)\rho_{\Lambda_l}(0)
\nonumber \\
+\sum_{\Lambda_L}\int_0^tdt_1\int_0^{t_1}dt_2\dots\int_0^{t_{L-1}}dt_L\, U_{\Lambda_0}(t,t_1)\mathcal{W}_{\Lambda_1}(t_1)U_{\Lambda_1}(t_1,t_2)\mathcal{W}_{\Lambda_2}(t_2)
\nonumber \\
\dots U_{\Lambda_{L-1}}(t_{L-1},t_L)\mathcal{W}_{\Lambda_L}(t_L)\rho_{\Lambda_L}(t_L)
\nonumber \\
+\frac{1}{S}\sum_{l=1}^L\sum_{\Lambda_{l-1}}\int_0^tdt_1\dots\int_0^{t_{l-1}}dt_l\, U_{\Lambda_0}(t,t_1)\mathcal{W}_{\Lambda_1}(t_1)U_{\Lambda_1}(t_1,t_2)\mathcal{W}_{\Lambda_2}(t_2)
\nonumber \\
\dots U_{\Lambda_{l-2}}(t_{l-2},t_{l-1})\mathcal{W}_{\Lambda_{l-1}}(t_{l-1})U_{\Lambda_{l-1}}(t_{l-1},t_l)\mathcal{V}_{\Lambda_{l-1}}(t_l)\rho_{\Lambda_{l-1}}(t_l),
\label{eq:Duhamel_rho}
\end{align}
and
\begin{align}
\sigma_{\Lambda_0}(t)=\sum_{l=0}^L\sum_{\Lambda_l}\int_0^tdt_1\int_0^{t_1}dt_2\dots\int_0^{t_{l-1}}dt_l\, U_{\Lambda_0}(t,t_1)\mathcal{W}_{\Lambda_1}(t_1)U_{\Lambda_1}(t_1,t_2)\mathcal{W}_{\Lambda_2}(t_2)
\nonumber \\
\dots U_{\Lambda_{l-1}}(t_{l-1},t_l)\mathcal{W}_{\Lambda_l}(t_l)U_{\Lambda_l}(t_l,0)\sigma_{\Lambda_l}(0)
\nonumber \\
+\sum_{\Lambda_L}\int_0^tdt_1\int_0^{t_1}dt_2\dots\int_0^{t_{L-1}}dt_L\, U_{\Lambda_0}(t,t_1)\mathcal{W}_{\Lambda_1}(t_1)U_{\Lambda_1}(t_1,t_2)\mathcal{W}_{\Lambda_2}(t_2)
\nonumber \\
\dots U_{\Lambda_{L-1}}(t_{L-1},t_L)\mathcal{W}_{\Lambda_L}(t_L)\sigma_{\Lambda_L}(t_L)
\nonumber \\
+\frac{1}{S}\sum_{l=1}^L\sum_{\Lambda_{l-1}}\sum_{(j,b)\in\Lambda_{l-1}}\int_0^tdt_1\dots\int_0^{t_{l-1}}dt_l\, U_{\Lambda_0}(t,t_1)\mathcal{W}_{\Lambda_1}(t_1)U_{\Lambda_1}(t_1,t_2)\mathcal{W}_{\Lambda_2}(t_2)
\nonumber \\
\dots U_{\Lambda_{l-2}}(t_{l-2},t_{l-1})\mathcal{W}_{\Lambda_{l-1}}(t_{l-1})U_{\Lambda_{l-1}}(t_{l-1},t_l)\mathcal{V}'_{\Lambda_{l-1},(jb')}(t_l)\sigma_{\Lambda_{l-1}\cup(jb')}(t_l).
\label{eq:Duhamel_sigma}
\end{align}
Here, $L$ is an arbitrary positive integer with $n+L<2S$.
The difference between Eqs.~(\ref{eq:Duhamel_rho}) and (\ref{eq:Duhamel_sigma}) is given by
\beq
\|\rho_{\Lambda_0}(t)-\sigma_{\Lambda_0}(t)\|_1^{\Lambda_0}\leq A_1+A_2+A_3+A_4,
\eeq
where
\begin{align}
A_1=\sum_{l=0}^L\sum_{\Lambda_l}\int_0^tdt_1\dots\int_0^{t_{l-1}}dt_l\, \|U_{\Lambda_0}(t,t_1)\mathcal{W}_{\Lambda_1}(t_1)U_{\Lambda_1}(t_1,t_2)\mathcal{W}_{\Lambda_2}(t_2)
\nonumber \\
\dots U_{\Lambda_{l-1}}(t_{l-1},t_l)\mathcal{W}_{\Lambda_l}(t_l)U_{\Lambda_l}(t_l,0)(\rho_{\Lambda_l}(0)-\sigma_{\Lambda_l}(0))\|_1^{\Lambda_0},
\end{align}
\begin{align}
A_2=\sum_{\Lambda_L}\int_0^tdt_1\int_0^{t_1}dt_2\dots\int_0^{t_{L-1}}dt_L\, \|U_{\Lambda_0}(t,t_1)\mathcal{W}_{\Lambda_1}(t_1)U_{\Lambda_1}(t_1,t_2)\mathcal{W}_{\Lambda_2}(t_2)
\nonumber \\
\dots U_{\Lambda_{L-1}}(t_{L-1},t_L)\mathcal{W}_{\Lambda_L}(t_L)(\rho_{\Lambda_L}(t_L)-\sigma_{\Lambda_L}(t_L))\|_1^{\Lambda_0},
\end{align}
\begin{align}
A_3=\frac{1}{S}\sum_{l=1}^L\sum_{\Lambda_{l-1}}\int_0^tdt_1\dots\int_0^{t_{l-1}}dt_l\, \|U_{\Lambda_0}(t,t_1)\mathcal{W}_{\Lambda_1}(t_1)U_{\Lambda_1}(t_1,t_2)\mathcal{W}_{\Lambda_2}(t_2)
\nonumber \\
\dots U_{\Lambda_{l-2}}(t_{l-2},t_{l-1})\mathcal{W}_{\Lambda_{l-1}}(t_{l-1})U_{\Lambda_{l-1}}(t_{l-1},t_l)\mathcal{V}_{\Lambda_{l-1}}(t_l)\rho_{\Lambda_{l-1}}(t_l)\|_1^{\Lambda_0},
\end{align}
and
\begin{align}
A_4=\frac{1}{S} \sum_{l=1}^L\sum_{\Lambda_{l-1}}\sum_{(j,b)\in\Lambda_{l-1}}\int_0^tdt_1\dots\int_0^{t_{l-1}}dt_l\, \|U_{\Lambda_0}(t,t_1)\mathcal{W}_{\Lambda_1}(t_1)U_{\Lambda_1}(t_1,t_2)\mathcal{W}_{\Lambda_2}(t_2)
\nonumber \\
\dots U_{\Lambda_{l-2}}(t_{l-2},t_{l-1})\mathcal{W}_{\Lambda_{l-1}}(t_{l-1})U_{\Lambda_{l-1}}(t_{l-1},t_l)\mathcal{V}'_{\Lambda_{l-1},(jb')}(t_l)\sigma_{\Lambda_{l-1}\cup(jb')}(t_l)\|_1^{\Lambda_0}.
\end{align}

We shall prove $\lim_{L\rightarrow\infty}\lim_{S\rightarrow\infty}A_i=0$ for $i=1,2,3,4$.
The following formulas are useful for doing that:
\begin{align}
\|U_{\Lambda_k}(t_k,t_{k+1})\hat{O}_{\Lambda_k}\|_1^{\Lambda_k}&=\|\hat{O}_{\Lambda_k}\|_1^{\Lambda_k},
\label{eq:formula_U}
\\
\sum_{(j_{k+1}b_{k+1})}\|\mathcal{W}_{\Lambda_{k+1}}(t_{k+1})\hat{O}_{\Lambda_{k+1}}\|_1^{\Lambda_k}&\leq J_0(n+k)\|\hat{O}_{\Lambda_{k+1}}\|_1^{\Lambda_{k+1}},
\label{eq:formula_W}
\\
\|\mathcal{V}_{\Lambda_k}(t_k)\hat{O}_{\Lambda_k}\|_1^{\Lambda_k}&\leq \frac{(n+k)^2}{4}J_0\|\hat{O}_{\Lambda_k}\|_1^{\Lambda_k},
\label{eq:formula_V}
\\
\|\mathcal{V}'_{\Lambda_k\cup(jb')}(t_k)\hat{O}_{\Lambda_k\cup(jb')}\|_1^{\Lambda_k}&\leq\frac{n+k}{4}J_0\|\hat{O}_{\Lambda_k\cup(jb')}\|_1^{\Lambda_k\cup(jb')},
\label{eq:formula_V'}
\end{align}
where $\hat{O}_{\Lambda_k}$ is an arbitrary operator acting on $\mathcal{H}_{\Lambda_k}$ and $J_0$ is defined in Eq.~(\ref{eq:interaction}).
Equation~(\ref{eq:formula_U}) is derived from the fact that $U_{\Lambda_k}(t_k,t_{k+1})$ is unitary.
Equation~(\ref{eq:formula_W}) is derived in the following way:
\begin{align}
\sum_{(j_{k+1}b_{k+1})}\|\mathcal{W}_{\Lambda_{k+1}}(t_{k+1})\hat{O}_{\Lambda_{k+1}}\|_1^{\Lambda_k}
&\leq \frac{1}{S}\sum_{(j_{k+1}b_{k+1})}\sum_{(ia)\in\Lambda_k}\left\|{\rm Tr}_{(j_{k+1}b_{k+1})}[\hat{V}_{(ia),(j_{k+1}b_{k+1})},\hat{O}_{\Lambda_{k+1}}]\right\|_1^{\Lambda_k}
\nonumber \\
&\leq \frac{1}{S}\sum_{(j_{k+1}b_{k+1})}\sum_{(ia)\in\Lambda_k}\left\|[\hat{V}_{(ia),(j_{k+1}b_{k+1})},\hat{O}_{\Lambda_{k+1}}]\right\|_1^{\Lambda_{k+1}}
\nonumber \\
&\leq \frac{2}{S}\sum_{(ia)\in\Lambda_k}\sum_{(j_{k+1}b_{k+1})}\|\hat{V}_{(ia),(j_{k+1}b_{k+1})}\|^{\Lambda_{k+1}}\|\hat{O}_{\Lambda_{k+1}}\|_1^{\Lambda_{k+1}}
\nonumber \\
&\leq \sum_{(ia)\in\Lambda_k}J_0\|\hat{O}_{\Lambda_{k+1}}\|_1^{\Lambda_{k+1}}
\nonumber \\
&=J_0(n+k)\|\hat{O}_{\Lambda_{k+1}}\|_1^{\Lambda_{k+1}},
\end{align}
where we have used the fact that the number of elements of $\Lambda_k$ is $n+k$ and the inequality
\beq
\sum_{j,b}\|\hat{V}_{(ia),(jb)}\|^{\Lambda_{k+1}}\leq \frac{1}{4}\sum_{j=1}^N\sum_{b=1}^{2S}\sum_{\alpha,\beta=x,y,z}|J_{ij}^{\alpha\beta}|
\leq \frac{S}{2}\sum_{j=1}^N\sum_{\alpha,\beta=x,y,z}|J_{ij}^{\alpha\beta}|\leq\frac{S}{2}J_0.
\eeq
Equation~(\ref{eq:formula_V}) is derived as follows:
\begin{align}
\|\mathcal{V}_{\Lambda_k}(t_k)\hat{O}_{\Lambda_k}\|_1^{\Lambda_k}&\leq\frac{1}{2}\sum_{(ia),(jb)\in\Lambda_k}\|[\hat{V}_{(ia),(jb)}(t_k),\hat{O}_{\Lambda_k}]\|_1^{\Lambda_k}
\nonumber \\
&\leq \sum_{(ia),(jb)\in\Lambda_k}\|\hat{V}_{(ia),(jb)}(t_k)\|^{\Lambda_k}\|\hat{O}_{\Lambda_k}\|_1^{\Lambda_k}
\nonumber \\
&\leq \frac{(n+k)^2}{4}J_0\|\hat{O}_{\Lambda_k}\|_1^{\Lambda_k}.
\end{align}
The last inequality is derived because
\begin{align}
\sum_{(ia),(jb)\in\Lambda_k}\|\hat{V}_{(ia),(jb)}(t)\|^{\Lambda_k}&\leq\frac{1}{4}\sum_{(ia)\in\Lambda_k}\sum_{(jb)\in\Lambda_k}\sum_{\alpha,\beta=x,y,z}|J_{ij}^{\alpha\beta}|
\nonumber \\
&\leq \frac{n+k}{4}\sum_{(ia)\in\Lambda_k}\sum_{j=1}^N\sum_{\alpha,\beta=x,y,z}|J_{ij}^{\alpha\beta}|
\nonumber \\
&\leq \frac{n+k}{4}\sum_{(ia)\in\Lambda_k}J_0
=\frac{(n+k)^2}{4}J_0.
\end{align}
Similarly, Eq.~(\ref{eq:formula_V'}) can be derived.

By using these formulas, we obtain
\begin{align}
A_1&\leq\sum_{l=0}^L\int_0^tdt_1\dots\int_0^{t_{l-1}}dt_l\, J_0^ln(n+1)\dots (n+l-1)\|\rho_{\Lambda_l}(0)-\sigma_{\Lambda_l}(0)\|_1^{\Lambda_l}
\nonumber \\
&=\sum_{l=0}^L\frac{(n+l-1)!}{(n-1)!l!}(J_0t)^l\|\rho_{\Lambda_l}(0)-\sigma_{\Lambda_l}(0)\|_1^{\Lambda_l}
\nonumber \\
&\leq \sum_{l=0}^L2^{n+l-1}(J_0t)^l\|\rho_{\Lambda_l}(0)-\sigma_{\Lambda_l}(0)\|_1^{\Lambda_l}
\nonumber \\
&=2^{n-1}\sum_{l=0}^L(2J_0t)^l\|\rho_{\Lambda_l}(0)-\sigma_{\Lambda_l}(0)\|_1^{\Lambda_l}.
\end{align}
Now we consider the time satisfying $0\leq t\leq t_0$ with $2J_0t_0=1/2$.
Then,
\beq
A_1\leq 2^{n-1}\sum_{l=0}^L2^{-l}\max_{k=1,2,\dots, L}\|\rho_{\Lambda_k}(0)-\sigma_{\Lambda_k}(0)\|_1^{\Lambda_k}
\leq 2^n\max_{k=1,2,\dots, L}\|\rho_{\Lambda_k}(0)-\sigma_{\Lambda_k}(0)\|_1^{\Lambda_k}.
\eeq
From the assumption of the initial state, for any fixed $k$,
\beq
\lim_{S\rightarrow\infty}\|\rho_{\Lambda_k}(0)-\sigma_{\Lambda_k}(0)\|_1^{\Lambda_k}=0.
\eeq
Therefore, $\lim_{S\rightarrow\infty}A_1=0$.

Next, we evaluate $A_2$ for $0\leq t\leq t_0$:
\begin{align}
A_2&\leq\int_0^tdt_1\dots\int_0^{t_{L-1}}dt_L\, J_0^Ln(n+1)\dots (n+L-1)\|\rho_{\Lambda_L}(t_L)-\sigma_{\Lambda_L}(t_L)\|_1^{\Lambda_L}
\nonumber \\
&\leq \frac{(n+L-1)!}{(n-1)!L!}(J_0t)^L\times 2
\nonumber \\
&\leq 2^n(2J_0t)^L\leq 2^{n-L},
\end{align}
where we have used $2J_0t\leq 2J_0t_0=1/2$.
If we take the limit of $L\rightarrow\infty$ after $S\rightarrow\infty$ (remember that $L$ is arbitrary as long as $n+L<S$), we have $A_2\rightarrow 0$.

Similarly, $A_3$ is evaluated as
\begin{align}
A_3&\leq \frac{1}{S}\sum_{l=1}^L\int_0^t dt_1\dots\int_0^{t_{l-1}}dt_l\, J_0^{l-1}n(n+1)\dots (n+l-2)\times \frac{(n+l-1)^2}{4}J_0
\nonumber \\
&\leq \frac{1}{S}\sum_{l=1}^L\frac{(n+l-1)!}{(n-1)!l!}(J_0t)^l\frac{n+l-1}{4}
\nonumber \\
&\leq\frac{1}{S}\sum_{l=1}^L2^{n-3}(2J_0t)^l(n+l-1)
\nonumber \\
&\leq\frac{2^{n-3}}{S}\sum_{l=1}^L2^{-l}(n+l-1)\leq \frac{2^{n-3}}{S}(n+1)
\end{align}
for $0\leq t\leq t_0$.
By taking the limit of $S\rightarrow\infty$, we obtain $A_3\rightarrow 0$.

As for $A_4$, we obtain for $0\leq t\leq t_0$,
\begin{align}
A_4&\leq \frac{1}{S}\sum_{l=1}^L\sum_{(jb)\in\Lambda_{l-1}}\int_0^tdt_1\dots\int_0^{t_{l-1}}dt_l\, J_0^{l-1}n(n+1)\dots (n+l-2)\times\frac{n+l-1}{4}J_0
\nonumber \\
&=\frac{1}{S}\sum_{l=1}^L(n+l-1)\frac{1}{4}\frac{(n+l-1)!}{(n-1)!l!}(J_0t)^l
\nonumber \\
&\leq \frac{2^{n-3}}{S}\sum_{l=1}^L(2J_0t)^l(n+l-1)\leq\frac{2^{n-3}}{S}(n+1).
\end{align}
Thus, in the limit of $S\rightarrow\infty$, we obtain $A_4\rightarrow 0$.
\end{widetext}

Up to now, we have shown $\lim_{L\rightarrow\infty}\lim_{S\rightarrow\infty}A_i=0$ for $i=1,2,3,4$ and $t\in[0,t_0]$.
It yields
\beq
\lim_{S\rightarrow\infty}\|\rho_{\Lambda_0}(t)-\sigma_{\Lambda_0}(t)\|_1^{\Lambda_0}=0
\label{eq:large_S}
\eeq
for any finite set $\Lambda_0$ and $t\in[0,t_0]$.
By doing the same evaluation for a new initial states $\rho_{\Lambda_0}(t_0)$ and $\sigma_{\Lambda_0}(t_0)$, we can show that Eq.~(\ref{eq:large_S}) holds for $t\in[t_0,2t_0]$.
In this way, by repeating the same argument, it is proved that Eq.~(\ref{eq:large_S}) holds for any fixed finite time $t\in[0,+\infty)$.
The proof is completed.

\section{Proof of Eq.~(\ref{eq:theorem2'})}
\label{appendix:theorem2}

Equation~(\ref{eq:theorem2'}) for an arbitrary initial state $|\psi(0)\>$ is equivalent to
\beq
s_M:=\left\|\hat{O}_X(MT)-\hat{O}_X^{(n)}(MT)\right\|\lesssim |X|T^{n+1}(MT)^{d+1}
\eeq
for large $MT$ with $M$ an integer, where
\beq
\hat{O}_X(MT)=e^{i\hat{H}_\mathrm{F}MT}\hat{O}_Xe^{-i\hat{H}_\mathrm{F}MT}
\eeq
and 
\beq
\hat{O}_X^{(n)}(MT)=e^{i\hat{H}_\mathrm{F}^{(n)}MT}\hat{O}_Xe^{-i\hat{H}_\mathrm{F}^{(n)}MT}.
\eeq
Without loss of generality, we can assume $\|\hat{O}_X\|=1$.
We can rewrite and evaluate $s_M$ as
\begin{align}
s_M=&\left\|e^{i\hat{H}_\mathrm{F}T}\hat{O}_X((M-1)T)e^{-i\hat{H}_\mathrm{F}T}\right.
\nonumber \\
&\left.-e^{i\hat{H}_\mathrm{F}^{(n)}T}\hat{O}_X^{(n)}((M-1)T)e^{i\hat{H}_\mathrm{F}^{(n)}T}\right\|
\nonumber \\
\leq&\left\|\hat{O}_X((M-1)T)-\hat{O}_X^{(n)}((M-1)T)\right\|
\nonumber \\
&+\left\|e^{i\hat{H}_\mathrm{F}T}\hat{O}_X((M-1)T)e^{-i\hat{H}_\mathrm{F}T}\right.
\nonumber \\
&\left.-e^{i\hat{H}_\mathrm{F}^{(n)}T}\hat{O}_X((M-1)T)e^{-i\hat{H}_\mathrm{F}^{(n)}T}\right\|,
\end{align}
and hence, we have
\begin{align}
s_M\leq s_{M-1}+\left\|e^{i\hat{H}_\mathrm{F}T}\hat{O}_X((M-1)T)e^{-i\hat{H}_\mathrm{F}T}\right.
\nonumber \\
\left.-e^{i\hat{H}_\mathrm{F}^{(n)}T}\hat{O}_X((M-1)T)e^{-i\hat{H}_\mathrm{F}^{(n)}T}\right\|.
\end{align}
Now we apply the Lieb-Robinson bound.
It is shown that $\hat{O}_X(t)$ can be well approximated by $\hat{O}_X(t;\ell)$ that is an operator acting only on the region
\beq
X_{\ell}:=\{i\in\Lambda: d(\{i\},X)\leq \ell\},
\eeq
that is, the set of all the sites whose distance from the region $X$ is at most $\ell$.
Explicitly, $\hat{O}_X(t;\ell)$ is defined by
\beq
\hat{O}_X(t;\ell)=\frac{1}{\mathrm{Tr}\,\hat{1}_{X_{\ell}^c}}\left[\mathrm{Tr}_{X_\ell^c}\hat{O}_X(t)\right]\otimes\hat{1}_{X_\ell^c},
\eeq
where $X_\ell^c$ is the complement of $X_\ell$, $\mathrm{Tr}_{X_\ell^c}$ is the partial trace over $\mathcal{H}_{X_\ell^c}$, and $\hat{1}_{X_\ell^c}$ is the identity operator acting onto $\mathcal{H}_{X_\ell^c}$.
It is noted that $\|\hat{O}_X(t;\ell)\|\leq\|\hat{O}_X\|=1$.
The Lieb-Robinson bound yields
\beq
\left\|\hat{O}_X(t)-\hat{O}_X(t;\ell)\right\|\leq ce^{-(\ell-vt)/\xi}|X|,
\label{eq:LR_2}
\eeq
where $c$, $v$, and $\xi$ are the constants appearing in Eq.~(\ref{eq:LR})~\cite{Bravi2006}.
Now we set
\beq
\ell=\ell_{M-1}:=v(M-1)T-\xi\ln f_T,
\eeq
where $0<f_T\leq 1$ is specified later, and use the notation $\hat{O}'_{X_{\ell_{M-1}}}:=\hat{O}_X((M-1)T;\ell_{M-1})$.
By using Eq.~(\ref{eq:LR_2}), we obtain
\begin{align}
&s_M\leq s_{M-1}+2|X|cf_T
\nonumber \\
&+\left\|e^{i\hat{H}_\mathrm{F}T}\hat{O}'_{X_{\ell_{M-1}}}e^{-i\hat{H}_\mathrm{F}T}
-e^{i\hat{H}_\mathrm{F}^{(n)}T}\hat{O}'_{X_{\ell_{M-1}}}e^{-i\hat{H}_\mathrm{F}^{(n)}T}\right\|.
\label{eq:sM_eq}
\end{align}
By using Eq.~(\ref{eq:single}), the last term of Eq.~(\ref{eq:sM_eq}) is evaluated as
\begin{align}
\left\|e^{i\hat{H}_\mathrm{F}T}\hat{O}'_{X_{\ell_{M-1}}}e^{-i\hat{H}_\mathrm{F}T}
-e^{i\hat{H}_\mathrm{F}^{(n)}T}\hat{O}'_{X_{\ell_{M-1}}}e^{-i\hat{H}_\mathrm{F}^{(n)}T}\right\|
\nonumber \\
\leq\alpha_n|X_{\ell_{M-1}}|T^{n+2}.
\end{align}
In a $d$-dimensional regular lattice, there exists a constant $a$ that depends only on the lattice geometry such that
\beq
|X_{\ell_{M-1}}|\leq a|X|\ell_{M-1}^d.
\eeq
Thus we have
\beq
s_M\leq s_{M-1}+2|X|cf_T+\alpha_na|X|T^{n+2}\ell_{M-1}^d.
\label{eq:rec}
\eeq
By repeatedly applying the inequality~(\ref{eq:rec}), we finally obtain
\beq
s_M\leq 2|X|cf_TM+\alpha_na|X|T^{n+2}\sum_{k=0}^{M-1}\ell_k^d.
\label{eq:SM_final}
\eeq
By replacing the summation by the integration, we have
\begin{align}
\sum_{k=0}^{M-1}\ell_k^d&\leq\frac{1}{T}\int_0^{MT}dt\,(vt-\xi\ln f_T)^d
\nonumber \\
&\leq\frac{1}{(d+1)vT}(vMT-\xi\ln f_T)^{d+1}.
\label{eq:integral}
\end{align}
Now we set
\beq
f_T=\left(\frac{T}{t_0}\right)^{n+2},
\label{eq:f_T}
\eeq
where $t_0$ is an arbitrary constant with the dimension of time satisfying $t_0>T$ and independent of $T$, e.g., $t_0=1/g$.
By substituting Eqs.~(\ref{eq:integral}) and (\ref{eq:f_T}) into Eq.~(\ref{eq:SM_final}), we obtain
\begin{align}
s_M\leq&\frac{2c}{t_0^{n+2}}|X|T^{n+1}MT
\nonumber \\
&+\frac{\alpha_nav^d}{d+1}|X|T^{n+1}\left[MT-\frac{\xi}{v}\ln(T/t_0)^{n+2}\right]^{d+1}.
\end{align}
For large $MT$, this behaves as
\beq
s_M\lesssim|X|T^{n+1}(MT)^{d+1},
\eeq
which is the desired result.

\bibliography{FM_classical.bib}

\end{document}